\documentclass[11pt,fleqn]{article}
\usepackage{amsfonts,amssymb,cite}
\usepackage{graphicx}
\usepackage{amsmath}
\usepackage{xcolor}
\def\noi{\noindent}

\newcommand{\Title}[1]{\noi {{\Large\bf #1}}\\[1ex]}

\newcommand{\Author}[2]{\noi{\bf #1}\\[2ex]\noi{\normalsize\it #2}\\}

\newcommand{\Abstract}[1]{\vskip 2mm \begin{center}
        \parbox{16.4cm}{\small\noi #1} \end{center}\medskip}
\newcommand{\foom}[1]{\protect\footnotemark[#1]}

\def\nqq{\hspace*{-2em}}

\def\lal{&&\nqq {}}

\def\beq{\begin{equation}}
\def\eeq{\end{equation}}
\def\bear{\begin{eqnarray}}
\def\bearr{\begin{eqnarray} \lal}
\def\ear{\end{eqnarray}}
\def\earn{\nonumber \end{eqnarray}}

\def\e{{\,\rm e}}

\newcommand{\MI}{$\mathfrak{M}_1$}
\newcommand{\MII}{$\mathfrak{M}_2$}

\newcommand{\MOO}{$\mathfrak{M}_{00}$}
\newcommand{\Eef}{E_{e\!f\!f}}
\newcommand{\Fig}[3]{%
\begin{center}
\parbox{8cm}{%
\refstepcounter{figure}\includegraphics[width=8cm,height=#2cm]{#1} \noindent Figure \thefigure:\quad
#3}\end{center}}
\begin{document}
\thispagestyle{empty}
\twocolumn[

\vspace{1cm}

\Title{Cosmological evolution of a statistical system of degenerate scalar-charged fermions with an asymmetric scalar doublet. I. Two-component system of assorted charges.\foom 1}

\Author{Yu. G. Ignat'ev$^1$, A.A. Agathonov$^2$ and D. Yu. Ignatyev$^1$}
    {$^1$Institute of Physics, Kazan Federal University, Kremlyovskaya str., 18, Kazan, 420008, Russia\\
    $^2$Lobachevsky Institute of Mathematics and Mechanics, Kazan Federal University, Kremlyovskaya str., 18, Kazan, 420008, Russia}


\Abstract
{Based on the mathematical model of a statistical system with scalar interaction of fermions formulated earlier, a cosmological model based on a two-component statistical system of scalar charged degenerate fermions interacting with an asymmetric scalar doublet, canonical and phantom scalar fields, is investigated. The asymptotic and limiting properties of the cosmological model are investigated, and it is shown that among all models there is a class of models with a finite lifetime. The asymptotic behavior of models near the corresponding singularities is investigated and numerical implementations of such models are constructed.
}
\bigskip

] 
\section*{Introduction}
Two-field models with the intersection of the barrier of the cosmological constant $w_{DE}=-1$ are known as quintom models and include one phantom scalar field and one standard (canonical) scalar field (see \cite{Feng} -- \cite{Vernov3}). These models have been intensively studied since 2005 due to the need to explain the observed cosmological acceleration of the Universe \cite{Odintsov}, \cite{Tsuj}
\begin{equation}\label{Omega}
\Omega=\frac{\ddot{a}a}{\dot{a}^2}\equiv 1+\frac{\dot{H}}{H^2}\equiv -\frac{1}{2}(1+3w),
\end{equation}
where $H$ is the Hubble parameter, $w$ is the barotropic coefficient
\begin{equation}\label{w}
w=\frac{p}{\varepsilon};
\end{equation}
$p$ and $\varepsilon$ are the total pressure and energy density of cosmological matter.

From a theoretical point of view, such models are very interesting and insufficiently studied. In this regard, we should note the work \cite{Lazkoz}, in which, based on the study of the dynamical system of the quintom, a counterexample of the typical behavior of the quintom, including attractors with $w\leqslant0$, was presented. In the works \cite{Yu_19_1} a qualitative and numerical analysis of quintom models was carried out under the assumption of non-negativity of the Hubble parameter ($H\geqslant0$). However, it was shown in \cite{Yu_20_1} that this assumption may contradict the complete system of Einstein's equations and the scalar field at $H\to0$. In the same paper, the concept of an Einstein-Higgs hypersurface was introduced, the topology of which largely determines the global properties of cosmological models based on an asymmetric scalar doublet. Thereby, a systematic study of the complete model of the cosmological evolution of an asymmetric scalar doublet (quintom) was carried out in \cite{YuKokh_TMF} using methods of qualitative theory of dynamical systems and numerical integration. A wide variety of behaviors of cosmological models depending on the topology of the Einstein - Higgs hypersurface was revealed, in particular, oscillatory modes characterized by alternating phases of expansion and compression were found. Note that in the recent independent papers \cite{Leon18} and \cite{ZhurCher} a qualitative analysis of cosmological models based on an asymmetric doublet with exponential potentials and chiral fields, respectively, was carried out.\footnote{We did not set the task to write any detailed review of works on two-field cosmological models based on canonical and phantom vacuum scalar fields. Such a review is contained, for example, in the work \cite{YuKokh_TMF} cited above.}

Note that we do not have any arguments in favor of the vacuum nature of cosmological scalar fields. These fields, like other physical fields, can be associated with corresponding charges, which, as usual, can correspond to some hypothetical fermions, $\zeta_a$, having scalar charges $q_a$. The interactions between these fermions and the quanta of the scalar field $\phi$ can be carried out in reactions of the type:
\begin{equation}\label{reactions}
\phi+\overline{\phi}\leftrightarrows \zeta+\overline{\zeta}; \quad \phi \leftrightarrows \zeta+\nu;\; \ldots
\end{equation}
This circumstance leads, first, to the necessity of constructing a selfconsistent mathematical model of the statistical system of scalarly charged fermions interacting with an asymmetric scalar doublet, and, secondly, to a comprehensive study of the cosmological evolution of such a statistical system and, thirdly, to the physical identification of $\zeta$-fermions, including in the composition of cosmic matter. In this sense, we believe it is promising to study statistical systems of completely degenerate $\zeta$-fermions, which, using the well-known processes of formation of Cooper pairs, can be transformed  into cold Bose condensate and create the basis of dark matter.

Scalar fields in general relativistic statistics and kinetics were introduced in the early 80's in the work \cite{Ignatev1}, and further, a non-minimal theory of scalar interaction was consistently developed on the basis of the concept of a fundamental scalar charge, both for canonical and phantom scalar fields \cite{Ignat_12_1_Iz}, \cite{Ignat_12_2_Iz}, \cite{Ignat_12_3_Iz}. In particular, in these works, some features of phantom fields were revealed, for example, features of interparticle interaction. Further, in the works \cite{Ignat_Dima14_2_GC}, \cite{Ignat_Agaf_Dima14_3_GC}, \cite{Ignat_15_1_GC}, a mathematical model of the statistical system of scalar charged particles was formulated, based on a microscopic description and the subsequent procedure for switching to kinetic and hydrodynamic models. Later, these studies were deepened to extend the theory of scalar fields, including phantom fields, to the sector of negative particle masses, degenerate Fermi systems, conformal-invariant interactions, etc. \cite{Ignat_Dima14_2_GC}, \cite{Ignat_Agaf_Dima14_3_GC}, \cite{Ignat_15_1_GC}, \cite{Ignat_Agaf15_2_GC}. The mathematical models of scalar fields constructed in this way were applied to the study of the cosmological evolution of systems of interacting particles and scalar fields, both of the canonical and phantom types \cite{Ignat_Mih15_2_Iz}, \cite{Ignat15_2}, \cite{Ignat_Sasha_G&G}. These studies revealed unique features of the cosmological evolution of plasma with interparticle phantom scalar interaction, such as the existence of giant bursts of cosmological acceleration, the presence of a plateau with constant acceleration, and other anomalies that sharply distinguish the behavior of cosmological models with a phantom scalar field from models with a canonical scalar field. The paper \cite{Ignat_Sasha_G&G} analyzes the types of behavior of cosmological models based on a scalar charged system of degenerate fermions with a quadratic interaction potential, and identifies 4 main types of behavior that allow, among other things, intermediate non-relativistic and ultra-relativistic expansion stages. Let us note that the intermediate non-relativistic stages of cosmological expansion are a necessary condition for the emergence of a large-scale structure of the Universe as a result of the development of gravitational instability. Also let us note that in models based on both vacuum scalar fields and vacuum fields with an ideal fluid, such steps do not occur spontaneously.

In \cite{Ignat20_1} a model of the cosmological evolution of a statistical system of degenerate scalar charged fermions interacting by means of a single scalar Higgs, canonical or phantom, field is formulated. This paper also provides examples of numerical models of such systems that radically differ in their behavior from the behavior of models based on vacuum scalar fields. On the basis of this model, in the works of one of the Authors \cite{Ignat_Stability_1, Ignat_Stability_2}, the theory of gravitational perturbations of a cosmological system of scalar charged fermions in the case of a scalar Higgs singlet (canonical or phantom) is developed and the stability of this cosmological model with respect to short-wave perturbations of the gravitational and scalar fields is investigated. In these papers, it is shown that the fermionic system is unstable in the early stages of cosmological expansion in the case of the canonical Higgs interaction and stable in the case of the phantom Higgs interaction.

Finally, in \cite{TMF_print21}, the macroscopic theory of statistical systems of scalar charged particles was justified at the microscopic level. In this paper, in particular, it was proved that the correct formula for the dynamic mass of scalar charged particles is \cite{Ignat_15_1_GC}
\begin{equation}\label{m_dyn}
m=m_0+\sum\limits_r q^r\Phi_r,
\end{equation}
where $m_0$ is some seed (bare) mass of the particle, $q^r$ is the scalar charge of the particle with respect to the $r$-th scalar field $\Phi_r$. In this case, the total mass of the particle is determined by the modulus of the dynamic mass
\[ m_*=\sqrt{m^2+p^2},\]
where $p$ is the momentum of the particle. In this paper, as in \cite{Ignat_15_1_GC}, it is shown that despite the fact that the dynamic mass \eqref{m_dyn} can take negative values, the macroscopic quantities determined by this mass give physically correct values.

Note that the question of the need for the seed mass $m_0$ in the formula \eqref{m_dyn} is quite subtle , because it is determined by the fundamental factor of the scalar fields $\Phi_r$ included in this formula. In the event that these fields are fundamental on a par with the gravitational field, and thus determine the masses of all particles, we must put $m_0\equiv0$, since otherwise the fundamental property of equality of the masses of particles and antiparticles will be violated. In the case where the scalar fields $\Phi_r$ are secondary and play an intermediate role in the structure and evolution of the universe, we can store the seed mass in the formula \eqref{m_dyn}, assuming that it will be removed at a deeper level of interactions.

In this connection, we note the following important property of statistical systems with scalar interaction \cite{TMF_print21}: in the case of a zero seed mass $m_0\equiv0$, such a system, unlike a system with electromagnetic interaction, admits a state with zero scalar fields. The interaction with scalar fields, including the appearance of the scalar charge density, is purely nonlinear. Thus, the vacuum state of such a system corresponds to zero potentials of scalar fields. Note that the complete Lagrangian of a cosmological system can, in principle, consist of a sum of the form
\begin{eqnarray}
L=\sum\limits_s L^{(s)}+L_G\equiv \nonumber\\
\sum\limits_{r_1} L^{(1)}_{r_1}+\sum\limits_{r_2}L^{(2)}_{r_2}+\ldots \sum\limits_{r_n}L^{(n)}_{r_n} +L_G,\nonumber
\end{eqnarray}
where $L_G$ is the Lagrangian of the gravitational field, and the Lagrangian $L^{(p)}$ and $L^{(q)}$ represent the Lagrangians of non-interacting physical subsystems. Thus, the real universe can be an object in the same Riemannian space with minimally interacting physical worlds (physical subsystems). Indirectly, we can estimate the number of such parallel worlds, given the fact that dark matter contributes to the total gravitational density of the universe of 22\%, while visible matter is only 4\% -- this number, taking into account our physical world, can be on the order of $5\div 6$ ($22:4=5\div6$). At the same time, the question of the interaction of fundamental scalar fields, both among themselves and with particles of physical subsystems, remains open.

This paper is devoted to the study of cosmological models based on the statistical system of scalar charged fully degenerate fermions with an asymmetric scalar doublet. Thus, our goal is, first, to extend the results of previous works to arbitrary values of the Hubble parameter based on the complete system of Einstein equations, second, to extend the results of previous works to the negative segment of the dynamic mass of fermions, and, third, to extend the theoretical model of the interaction of fermions with scalar fields to the model of the asymmetric Higgs doublet.

Finally, in \cite{TMF_print21}, two simplest models for the interaction of fermions with an asymmetric scalar doublet were proposed: in the first model, such interaction is carried out by two types of assorted fermions, one of which is the source of the canonical scalar field, and another -- the phantom field (model \MI); in the second model, there is one class of fermions with a paired charge -- the canonical and phantom (model \MII). A qualitative analysis of the dynamic systems of the corresponding models was also carried out there. In this paper, we firstly investigate the first model (\MI) of a two-component statistical system with asymmetric scalar interaction and compare it with the results of previous studies. In this part of the paper, due to the large number of parameters of the model (11), we will limit ourselves to the case of zero seed mass of $\zeta$ - fermions and to the demonstration of the most typical cases of the behavior of the cosmological model.

\section{Mathematical model of the cosmological plasma of charged\\ fermions with the quintom interaction}
Consider the space-plane model of the Friedman universe\footnote{In this paper, we use a metric with the signature $(---+)$, the Ricci tensor is defined by the convolution of the first and third indices of the curvature tensor (see, for example, \cite{Land_Field})}:
\begin{equation}\label{ds}
ds^2=dt^2-a^2(t)(dx^2+dy^2+dz^2).
\end{equation}
Consider a statistical system consisting of $n$ varieties of degenerate scalar-charged fermions with scalar charges $q_{(a)}^r$ with respect to $N$ scalar fields $\Phi_r$. Let further
\begin{equation}\label{m_*(pm)}
m_{(a)}=m^0_{(a)}+\sum\limits_r q_{(a)}^r\Phi_r
\end{equation}
be the dynamic masses of these fermions and $L_s$ is the Lagrange function of non-interacting scalar Higgs fields $L_s=\sum\limits_r L_{(r)}$
\begin{eqnarray} \label{Ls}
L_s=\frac{1}{16\pi}\sum\limits_r(e_r g^{ik} \Phi_{r,i} \Phi _{r,k} -2V_r(\Phi_r)),
\end{eqnarray}
\begin{eqnarray}
\label{Higgs}
V_r(\Phi_r)=-\frac{\alpha_r}{4} \left(\Phi_r^{2} -\frac{m_r^{2} }{\alpha_m}\right)^{2}
\end{eqnarray}
-- the potential energy of the corresponding scalar fields, $\alpha_r$ -- their selfaction constants, $m_r$ -- masses of their quanta , $e_r= \pm 1$ - indicators (the ``+'' sign corresponds to canonical scalar fields, the ``-'' sign corresponds to phantom fields).

Further, the energy-momentum tensor of scalar fields with respect to the Lagrange function \eqref{Ls} is:
\begin{eqnarray}\label{T_s}
\!\!T^i_{(s)k}=\!\frac{1}{16\pi }\sum\limits_r\bigl(2\e_r\Phi^{,i}_{r} \Phi _{r,k} -e_r\delta^i_k\Phi _{r,j} \Phi _r^{,j}
 \nonumber\\+2V_r(\Phi_r)\delta^i_k \bigr),
\end{eqnarray}
and the energy-momentum tensor of an equilibrium statistical system is equal to:
\begin{equation}\label{T_p}
T^i_{(p)k}=(\varepsilon_p+p_p)u^i u_k-\delta^i_k p,
\end{equation}
where $u^i$ is the vector of the macroscopic velocity of the statistical system, $\varepsilon_p$ and $p_p$ are its energy density and pressure. Einstein's equations for the system ``scalar fields+particles'' have the form:
\begin{equation}\label{Eq_Einst_G}
R^i_k-\frac{1}{2}\delta^i_k R=8\pi T^i_k+ \delta^i_k \Lambda_0,
\end{equation}
where
\[T^i_k=T^i_{(s)k}+T^i_{(p)k},\]
$\Lambda_0$ is the seed value of the cosmological constant associated with its observed value $\Lambda$ by the relation:
\begin{equation}\label{lambda0->Lambda}
\Lambda=\Lambda_0-\frac{1}{4}\sum\limits_r \frac{m^4_r}{\alpha_r}.
\end{equation}
\subsection{General relations for models of degenerate fermions with scalar interaction}
 It can be proved\footnote{For a one-component system of degenerate fermions, see \cite{Ignat_Agaf_Dima14_3_GC}, \cite{Ignat15_2}, and for a multi-component system, see \cite{TMF_print21}}, that a strict consequence of the general relativistic kinetic theory for statistical systems of completely degenerate fermions is the Fermi momentum conservation law $\pi_{(a)}$ of each component
 \begin{equation}\label{ap}
 a(t)\pi_{(a)}(t)=\mathrm{Const}.
 \end{equation}
Assuming further for certainty $a (0)=1$ (see note on p. \pageref{attan}) and
\begin{equation}\label{a-xi}
\xi=\ln a;\quad \xi\in(-\infty,+\infty); \quad \xi(0)=0,
\end{equation}
we introduce dimensionless functions
\begin{equation}\label{psi0}
\psi_{(a)}=\frac{\pi^0_{(a)} \mathrm{e}^{-\xi}}{|m_{(a)}|}, \quad (\pi^0_{(a)}=\pi_{(a)}(0)),
\end{equation}
equal to the ratio of the Fermi momentum $\pi_{(a)}$ to the total energy of the fermion, as well as the functions $F_1(\psi)$ and $F_2(\psi)$:
\begin{equation}\label{F_1}
F_1(\psi)=\psi\sqrt{1+\psi^2}-\ln(\psi+\sqrt{1+\psi^2});
\end{equation}
\begin{equation}
\label{F_2}
F_2(\psi)=\psi\sqrt{1+\psi^2}(1+2\psi^2)-\ln(\psi+\sqrt{1+\psi^2}),
\end{equation}
with the help of which we define the macroscopic scalars of the statistical system:
density of the number of particles of the sort $a$ --
\begin{equation}\label{2_3}
n^{(a)}=\frac{1}{\pi^2}\pi_{(a)}^3;
\end{equation}
pressure --
\begin{equation}\label{2_3a}
{\displaystyle
\begin{array}{l}
\varepsilon_p = {\displaystyle\sum\limits_a\frac{m_{(a)}^4}{8\pi^2}}F_2(\psi_{(a)});
\end{array}}
\end{equation}
energy density of the fermion system --
\begin{equation}\label{2_3b}{\displaystyle
\begin{array}{l}
p_p ={\displaystyle\sum\limits_a\frac{m_{(a)}^4}{24\pi^2}}(F_2(\psi_{(a)})-4F_1(\psi_{(a)}))
\end{array}}
\end{equation}
and the density of the scalar charge of the fermion system with respect to the scalar field $\Phi_r$ --
\begin{equation}\label{2_3c}{\displaystyle
\begin{array}{l}
\sigma_r={\displaystyle \sum\limits_a q^r_{(a)}\frac{m_{(a)}^3}{2\pi^2}}F_1(\psi_{(a)}).
\end{array}}
\end{equation}
Note also the useful relation
\begin{equation}\label{E_P_f}
\varepsilon_p+p_p\equiv \frac{1}{3\pi^2}\sum\limits_a m^4_{(a)}\psi_{(a)}^3\sqrt{1+\psi_{(a)}^2}.
\end{equation}

Thus, in a cosmological model consisting of a system of degenerate scalar-charged fermions and scalar fields, all macroscopic scalars are defined by explicit algebraic functions of the scalar potentials $\Phi_1(t),\ldots,\Phi_N(t)$.

A complete closed normal autonomous system of ordinary differential equations describing a cosmological model based on a statistical system of fully degenerate scalar charged fermions has the form (see \cite{Ignat20_1}, \cite{TMF_print21}):
\begin{eqnarray}
\label{dot(xi)}
\dot{\xi}=H;\\
\label{dot(H)}
\dot{H}=-\sum\limits_r\frac{ e_r Z^2_r}{2}-\sum\limits_a \frac{4m_{(a)}^2\psi_{(a)}^3}{3\pi}\sqrt{1+\psi_{(a)}^2};\\
\label{dot(Phi)}
\dot{\Phi_r}=Z_r, \qquad (r=\overline{1,N});\\
\label{dot(Z)}
e_r\dot{Z}_r=-e_r 3HZ_r- m^2_r\Phi_r+\alpha_r\Phi^3_r-8\pi\sigma_r(t).
\end{eqnarray}

\textbf{\label{attan}Remark}. Let us note, first, that since the dynamical system \eqref{dot(xi)} -- \eqref{dot(Z)} is \emph{an autonomous system of differential equations}, it is invariant under time translations $t\to t_0+t$. Therefore, when studying this system, any value can be chosen as the initial moment of time $t_0$, including $t_0=0$. Of course, this meaning has nothing to do with the ``beginning of the universe'' in the general case. Due to this circumstance, the time $t$ can be continued into the negative range of values, if this is not prevented by a possible singularity.

A strict consequence of the dynamic system \eqref{dot(xi)} -- \eqref{dot(Z)} is the total energy integral $3H^2-8\pi\Eef=\mathrm{Const}$, whose partial zero value corresponds to the Einstein equation $^4_4$
\begin{eqnarray}\label{Surf_EinstNn}
3H^2-\Lambda-\sum\limits_r \left(\frac{e_rZ^2}{2}-\frac{m^2_r\Phi^2_r}{2}+\frac{\alpha_r\Phi^4_r}{4}\right)\nonumber\\
-\frac{1}{\pi}\sum\limits_a m^4_{(a)}F_2(\psi_{(a)} \equiv 3H^2-8\pi \Eef =0.
\end{eqnarray}
The equations \eqref{Surf_EinstNn} defines a certain hypersurface $\Sigma_E$ in the $2 N+2$ - dimensional phase space of the dynamical system \eqref{dot(xi)} -- \eqref{dot(Z)}. This hypersurface will henceforth be called the Einstein hypersurface. The points of the phase space $\mathbb{R}_{2 N+2}$ where the effective energy $\Eef$ is negative are not available for the dynamical system. The inaccessible region is separated from the accessible region of the phase space by a hypersurface of zero effective energy $S_{E}\subset\mathbb{R}_{2N+2}$, which is a cylinder with the $OH$ axis:
\begin{eqnarray}\label{S_ENn}
8\pi\Eef\equiv \Lambda+\frac{1}{\pi}\sum\limits_a m^4_{(a)}F_2(\psi_{(a)}\nonumber\\
+\sum\limits_r \left(\frac{e_rZ^2}{2}+\frac{m^2_r\Phi^2_r}{2}-\frac{\alpha_r\Phi^4_r}{4}\right)
=0,
\end{eqnarray}
moreover, the hypersurface of zero effective energy \eqref{S_ENn} touches the Einstein hypersurface \eqref{Surf_EinstNn} in the hyperplane
$H=0$:
\begin{equation}\label{H=0}
\Sigma_E \cap S_E =H=0.
\end{equation}
The fact that the Einstein equation \eqref{Surf_EinstNn} is a particular first integral of the dynamical system \eqref{dot(xi)} -- \eqref{dot(Z)} allows us to use it to determine the initial value of the Hubble parameter $H_0=H(0)$, which we will do in the future. The equation \eqref{Surf_EinstNn} is square with respect to $H(t)$, so it has two symmetric roots $\pm H(t)$, where the positive root corresponds to the expansion of the universe, and the negative root corresponds to the contraction.

\subsection{The main relations of the cosmological model with the quintom interaction of fermions (the \MI model)}
Next, $L_s$ is the Lagrange function of interacting scalar fields of canonical ($\Phi$) and phantom ($\varphi$)
\begin{eqnarray} \label{Ls}
L_s= \frac{1}{16\pi}(g^{ik} \Phi_{,i} \Phi _{,k} -2V(\Phi))\nonumber\\
+\frac{1}{16\pi}(- g^{ik} \varphi_{,i} \varphi _{,k} -2\mathcal{V}(\varphi)),
\end{eqnarray}
where
\begin{eqnarray}
\label{Higgs}
\!\!\!\!V(\Phi)\!=\!-\frac{\alpha}{4} \left(\Phi^{2}\! -\frac{m^{2} }{\alpha}\right)^{2}\!;
\mathcal{V}(\varphi)\!=\!-\frac{\beta}{4} \left(\varphi^2\! -\frac{\mathfrak{m}^2}{\beta}\right)^{2}\nonumber\
\end{eqnarray}
-- the potential energy of the corresponding scalar fields, $\alpha$ and $\beta$ -- the constants of their selfaction, $m$ and $\mathfrak{m}$ -- masses of their quanta. As a carrier of quintom charges, we consider a two - component degenerate system of fermions, in which the carriers of the canonical charge $\zeta_c$ -- fermions have a seed mass $m^0_c$, the canonical charge $e$ and the initial Fermi momentum $\pi^0_c$, and the carriers of the phantom charge $\zeta_f$ -- fermions have a seed mass $m^0_f$, the phantom charge $\epsilon$ and the initial Fermi momentum $\pi^0_f$. We write out a complete normal system of Einstein equations and scalar fields $\Phi (t)$ and $\varphi (t)$ for this two-component system of scalar charged degenerate fermions\cite{TMF_print21}\footnote{For a scalar singlet, this system is obtained in \cite{Ignat20_1}},\footnote{For simplicity, we have written out a system of dynamic equations for the case of the symmetric mass formula \eqref{m_*(pm)}, when $m^0_{(a)}\equiv0$, although examples with $m^0_{(a)}>0$ will also be considered below}.
In an obviously nonsingular form, the normal system of ordinary differential equations of the model under study has the form:
\begin{eqnarray}
\label{dxi/dt-dPhi_Phi}
\dot{\xi}=H;\qquad \dot{\Phi}=Z;\qquad \dot{\varphi}=z;\\
\label{dH/dt_M1}
\dot{H}=-\frac{Z^2}{2}+\frac{z^2}{2}-\frac{4\mathrm{e}^{-3\xi}}{3\pi}\times\nonumber\\
\biggl(\pi_c^3\sqrt{\pi_c^2\mathrm{e}^{-2\xi}+e^2\Phi^2}+\pi_f^3\sqrt{\pi_f^2\mathrm{e}^{-2\xi}+\epsilon^2\varphi^2}\biggr);\\
\label{dZ/dt_M1}
\dot{Z}=-3HZ-m^2\Phi+\alpha\Phi^3\nonumber\\
-\frac{4e^2\pi_c\mathrm{e}^{-\xi}}{\pi}\Phi\sqrt{\pi^2_c \mathrm{e}^{-2\xi}+e^2\Phi^2}\nonumber\\
+\frac{4e^4}{\pi}\Phi^3\ln\biggl(\frac{\pi_c\mathrm{e}^{-\xi}+\sqrt{\pi^2_c \mathrm{e}^{-2\xi}+e^2\Phi^2}}{|e\Phi|} \biggr);\\
\label{dzZ/dt_M1}
\dot{z}=-3Hz+\mathfrak{m}^2\varphi-\beta\varphi^3\nonumber\\
+\frac{4\epsilon^2\pi_f\mathrm{e}^{-\xi}}{\pi}\varphi\sqrt{\pi^2_f \mathrm{e}^{-2\xi}+\epsilon^2\varphi^2}-\nonumber\\
\frac{4\epsilon^4}{\pi}\varphi^3\ln\biggl(\frac{\pi_f\mathrm{e}^{-\xi}+\sqrt{\pi^2_f \mathrm{e}^{-2\xi}+\epsilon^2\varphi^2}}{|\epsilon\varphi|} \biggr),
\end{eqnarray}
where
\begin{equation}\label{psi-psi}
\psi_c=\frac{\pi^0_c}{|e\Phi|}\mathrm{e}^{-\xi}; \quad \psi_f=\frac{\pi^0_f}{|\epsilon\varphi|}\mathrm{e}^{-\xi}.
\end{equation}
The system of equations \eqref{dxi/dt-dPhi_Phi} -- \eqref{dzZ/dt_M1} has as its first integral the total energy integral \cite{TMF_print21}, which can be used to determine the initial value of the function $H(t)$
\begin{eqnarray}
\frac{Z^2}{2}+\frac{z^2}{2}-\frac{m^2\Phi^2}{2}+\frac{\alpha\Phi^4}{4}-\frac{\mathfrak{m}^2\varphi^2}{2}+\frac{\beta\varphi^4}{4} \nonumber\\
-\frac{e^{-\xi}}{\pi}\biggl(\pi_c\sqrt{\pi_c^2\mathrm{e}^{-2\xi}+e^2\Phi^2}\bigl(2\pi^2_c\mathrm{e}^{-2\xi}+e^2\Phi^2\bigr)\nonumber\\
+\pi_f\sqrt{\pi_f^2\mathrm{e}^{-2\xi}+\epsilon^2\varphi^2}\bigl(2\pi^2_f \mathrm{e}^{-2\xi}+\epsilon^2\varphi^2\bigr)\biggr)\nonumber
\end{eqnarray}
\begin{eqnarray}\label{SurfEinst_M1}
+\frac{e^4\Phi^4}{\pi}\ln\biggl(\frac{\pi_c\mathrm{e}^{-\xi}+\sqrt{\pi^2_c \mathrm{e}^{-2\xi}+e^2\Phi^2}}{|e\Phi|} \biggr)\nonumber\\
+\frac{\epsilon^4\varphi^4}{\pi}\ln\biggl(\frac{\pi_f\mathrm{e}^{-\xi}+\sqrt{\pi^2_f \mathrm{e}^{-2\xi}+\epsilon^2\varphi^2}}{|\epsilon\varphi|} \biggr)\nonumber\\
+3H^2-\Lambda-=0.
\end{eqnarray}
In the future, we will call this model \MI.
\subsection{Limit and asymptotic properties of the \MI model\label{ssM1}}
Note, first, that in the absence of fermions ($\pi_c=\pi_f=0$), the system of equations \eqref{dxi/dt-dPhi_Phi} -- \eqref{dzZ/dt_M1}, \eqref{SurfEinst_M1} continuously proceeds to the system of equations for the vacuum Higgs doublet (see \cite{YuKokh_TMF}):
\begin{eqnarray}
\label{dH/dt_Vac}
\dot{H}=-\frac{Z^2}{2}+\frac{z^2}{2};\\
\label{dZ/dt_Vac}
\dot{Z}=-3HZ-m^2\Phi+\alpha\Phi^3;\\
\label{dz/dt_Vac}
\dot{z}=-3Hz+\mathfrak{m}^2\varphi-\beta\varphi^3;\\
\label{SurfEinst_Vac}
3H^2-\Lambda-\frac{Z^2}{2}+\frac{z^2}{2}-\frac{m^2\Phi^2}{2}+\frac{\alpha\Phi^4}{4} \nonumber\\
-\frac{\mathfrak{m}^2\varphi^2}{2}+\frac{\beta\varphi^4}{4}=0.
\end{eqnarray}

Second, note that the beginning of the universe (the cosmological singularity $a=0$) corresponds to $\xi\to -\infty$, and the infinite future $a\to\infty$ -- $\xi\to+\infty$ (if the model allows such a state). One can easily see that for $\xi\to+\infty$, the system of equations \eqref{dxi/dt-dPhi_Phi} -- \eqref{dzZ/dt_M1} also asymptotically tends to the system of equations for the vacuum Higgs doublet \eqref{dH/dt_Vac} -- \eqref{SurfEinst_Vac}. Therefore, if the cosmological model admits the state $\xi\to+\infty$\footnote{This condition is not allowed in all cases of model parameters, as can be seen in the examples below.}. then, to study the evolution of the cosmological model at the later stages, we can apply the results of qualitative and numerical analysis of the vacuum model of the asymmetric scalar doublet \cite{YuKokh_TMF}.

Third, if $\pi_c=0, \varphi=0$ or $\pi_f=0,\Phi=0$, the model becomes a model of a single-component scalar charged Fermi system with the corresponding scalar singlet \cite{Ignat20_1}.

Fourth, for zero charges of $\zeta$ - fermions, the system of equations \eqref{dxi/dt-dPhi_Phi} -- \eqref{dzZ/dt_M1}, \eqref{SurfEinst_M1} continuously segues to a system of equations for a cosmological model based on a vacuum asymmetric scalar Higgs doublet and a neutral two-component Fermi liquid.

Finally, we investigate the behavior of the cosmological model near the cosmological singularity $\xi\to - \infty$. From the equations \eqref{dxi/dt-dPhi_Phi} -- \eqref{dzZ/dt_M1}, it follows that such a state is always possible for $\pi_c,\pi_f\not\equiv 0$. In this case, for $\xi\to-\infty$, $H\to\pm\infty$ $\mathrm{e}^{\xi}|\Phi|\ll 1$, $|\varphi|<\infty$, the system of equations \eqref{dxi/dt-dPhi_Phi} -- \eqref{dzZ/dt_M1} reduces to the following:
\begin{eqnarray}
\label{dxi/dt-dPhi_Phi_8}
\dot{\xi}=H;\qquad \dot{\Phi}=Z;\qquad \dot{\varphi}=z;\\
\label{dH/dt_M1_8}
\dot{H}=-\frac{4\mathrm{e}^{-4\xi}}{3\pi}(\pi^4_c+\pi^4_f); \\
\label{dZ/dt_M1_8}
\dot{Z}=-3HZ-\frac{4e^2\pi^2_c\mathrm{e}^{-2\xi}}{\pi}\Phi \ ;\\
\label{dz/dt_M1_8}
\dot{z}=-3Hz+\frac{4\epsilon^2\pi^2_f\mathrm{e}^{-2\xi}}{\pi}\varphi.
\end{eqnarray}

Replacing the variable in the equation \eqref{dH/dt_M1} $d/dt=H d/d\xi$, we find its solution;
\begin{eqnarray}\label{H_xi-8}
H=\pm\sqrt{\frac{2}{3}(\pi^4_c+\pi^4_f)}\mathrm{e}^{-2\xi},
\end{eqnarray}
where the sign $+$ corresponds to the exit from the singularity, $-$ corresponds to the entrance to it. Substituting in \eqref{H_xi-8} $H$ from the equation \eqref{dxi/dt-dPhi_Phi_8}, we get the differential equation with respect to $\xi(t)$:
\begin{eqnarray}
\dot{\xi}=\pm \sqrt{\frac{2}{3}(\pi^4_c+\pi^4_f)}\mathrm{e}^{-2\xi}\Rightarrow\nonumber\\
d\mathrm{e}^{2\xi}=\pm \sqrt{\frac{8}{3}(\pi^4_c+\pi^4_f)}dt,\nonumber
\end{eqnarray}
from where, taking into account the definition of $\xi(t)$ \eqref{a-xi} and the condition for the existence of a singularity at the point $t_0:\ \xi(t_0)=-\infty$ $\Rightarrow a(t_0)=0$, we get the asymptotics of the proximity of the singularity:
\begin{eqnarray}\label{a(t_0)}
\mathrm{e}^\xi\equiv a(t)\propto \displaystyle\biggl(\frac{8}{3}(\pi^4_c+\pi^4_f)\biggr)^{\frac{1}{4}}\sqrt{|t-t_0|}.
\end{eqnarray}
Thus, near the singularity $t\to t_0$, the scale factor and the Hubble parameter have the following asymptotics:
\begin{equation}\label{a(t0),H(t0)}
\left.a(t)\right|_{t\to t_0}\propto \sqrt{|t-t_0|}; \quad \left.H(t)\right|_{t\to t_0}\propto \frac{1}{t-t_0}.
\end{equation}
Thus, for $t<t_0$ $H<0$ (compression), for $t>t_0$ $H>0$ (expansion).

Finally, calculating the invariant cosmological acceleration $\Omega$ \eqref{Omega} with \eqref{dH/dt_M1_8}, \eqref{H_xi-8} and \eqref{a(t_0)} we obtain near the singularity:
\begin{equation}\label{Omega_8}
\left.\Omega(t)\right|_{t\to t_0}\propto -1,
\end{equation}
which corresponds, as we know, to the ultrarelativistic equation of state $w=\frac{1}{3}$.

Substituting, finally, \eqref{H_xi-8}, \eqref{a(t_0)} into \eqref{dZ/dt_M1_8} and \eqref{dz/dt_M1_8}, we finally find the asymptotics for scalar potentials near the cosmological singularity $\xi(t_0)=-\infty$ ($a(t_0)=0$):%
\begin{eqnarray}\label{Phi(t)_M1_-8}
\displaystyle \Phi\simeq & \tilde{C}_+(t-t_0)^{\frac{3}{4}(1-\theta_\mp)}
+\tilde{C}_-(t-t_0)^{\frac{3}{4}(1+\theta_\mp)};\nonumber\\
\label{varphi_M1(t)_-8}
\displaystyle \varphi\simeq &\tilde{c}_+(t-t_0)^{\frac{3}{4}(1-\vartheta_\mp)}
+\tilde{c}_-(t-t_0)^{\frac{3}{4}(1+\vartheta_\mp)},
\end{eqnarray}
where
\begin{equation}\label{theta-vartheta}
\theta_\mp=\sqrt{1\mp\frac{\rho^2}{9}};\quad \vartheta_\mp=\sqrt{1\mp\frac{\varrho^2}{9}}.
\end{equation}

\section{Numerical simulation}
\subsection{Relation of the \MI model to the previously investigated \MOO model}
We will call the cosmological model with a one-component system of scalar charged fermions with a scalar singlet with a quadratic interaction potential, a nonnegative dynamic mass, and a nonnegative Hubble parameter $H\geqslant0$, studied in \cite{Ignat15_2}, \cite{Ignat_Sasha_G&G}, the \MOO model. To move to this model, in the system of equations \eqref{dxi/dt-dPhi_Phi} -- \eqref{dzZ/dt_M1}, \eqref{SurfEinst_M1}, first, it is necessary to put zero potentials and their derivatives of one of the scalar fields, second, to put equal to zero the Fermi momentum of the corresponding $\zeta$-fermions, and third, to replace the expression for the dynamic mass in the scalar charge density by its modulus $m\to|m|$, finally, fourth, replace in the system of dynamic equations \eqref{dxi/dt-dPhi_Phi} -- \eqref{dzZ/dt_M1} equation \eqref{dH/dt_M1} equation \eqref{SurfEinst_M1}, from which to determine the non-negative root $H_+$, and finally put $\alpha=\beta=0$.

Next, to shorten the letter, we will set a set of fundamental parameters of the \MI model using an ordered list
\[\mathbf{P}=[[\alpha,m,e,m_c,\pi_c],[\beta,\mu,\epsilon,m_f,\pi_f],\Lambda]\]
and the initial conditions are an ordered list
\[\mathbf{I}=[\Phi_0,Z_0,\varphi_0,z_0,\chi],\]
where $\chi=\pm1$, and the value of $\chi=+1$ corresponds to the non-negative initial value of the Hubble parameter $H_0=H_+ \geqslant0$, and the value of $\chi=-1$ corresponds to the negative initial value of the Hubble parameter $H_0=H_-<0$. At the same time, using the autonomy of the dynamical system, we everywhere assume $\xi(0)=0$. Thus, the \MI model is defined by 11 fundamental parameters and 5 initial conditions. The \MOO model is defined by 5 fundamental parameters $\mathbf{P}_{00}=[\mu,\mathbf{e}, q, m,\pi_0,\Lambda]$ and one indicator $\mathbf{e}=\pm1$, and $\mathbf{e}=+1$ corresponds to the canonical field, $\mathbf{e}=-1$ - to the phantom field. The initial conditions for this model are set by a list of two elements $\mathbf{I}_{00}=[\Phi_0, Z_0]$.

\subsection{Comparison of the behavior of the \MI and \MOO models}
The main difference between the full \MI model and the incomplete \MOO model is the appearance of the vibrational nature of the metric functions $ \ xi(t)$, $a(t)$, $H (t)$ in comparison with the monotonic nature of these functions in the \MOO model. In Fig. \ref{xi(t)_MM0M00} -- \ref{H(t)_MM0M00} the dependence of the evolution of these values on the mathematical model is demonstrated.

\Fig{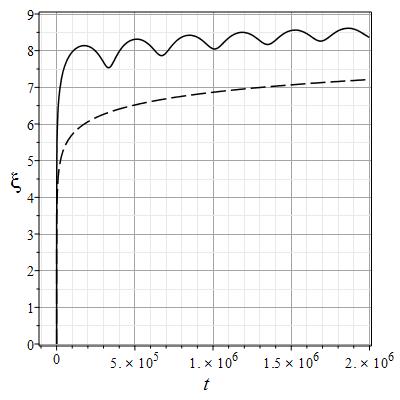}{8}{\label{xi(t)_MM0M00} Evolution of the function $\xi(t)=\ln(a(t))$ in various models with a single phantom field: solid line -- model \MI, dashed -- model \MOO. Everywhere $\beta=0,\mu=5\cdot10^{-8}$.}

It should be noted that the monotonic increase of the metric function $\xi(t)$ in the \MOO model is precisely provided by the condition of non-negativity of the Hubble parameter.

\Fig{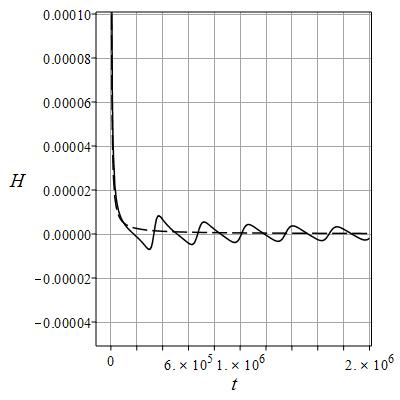}{8}{\label{H(t)_MM0M00} Evolution of the Hubble parameter $H(t)$ in various models with a single phantom field: solid line -- model \ MI, dashed -- model \MOO for the parameters Fig. \ref{xi(t)_MM0M00}.}

\section{Example of standard behavior of the cosmological model $\Lambda>0$}
Let us consider an example of the ``standard'' behavior of a cosmological model, when it has all the main features of the corresponding model with vacuum Higgs fields
\[\mathbf{P}_0=[[1,1,1,0.1,0.2],[1,1,0.5,0.1,0.2],0.1];\]
\[\mathbf{I}_0=[0.1,0.1,0.1,0.1,1].\]
In Fig. \ref{xi0} -- \ref{H0_1} shows the cosmological evolution of metric functions for this model. As can be seen from these figures, this model describes the transition from the compression phase $H_-\approx-0.34$, $\omega=1$ to the expansion stage $H_+\approx +0.34$,$\Omega=1$. This case is quite typical for cosmological models with a vacuum asymmetric doublet \cite{YuKokh_TMF}.

Next, in Fig. \ref{FZ0} -- \ref{f_z0} the corresponding phase trajectories of scalar fields are presented. These trajectories are also quite typical for the model with an asymmetric vacuum Higgs doublet, when the trajectory of the canonical field is wound on the zero focus $[0,0]$, and the trajectory of the phantom field is wound on the right focus $[1,0]$.

\Fig{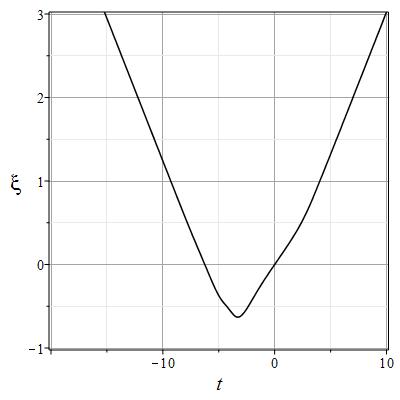}{8}{\label{xi0} Evolution of the scale function $\xi (t)$ in the \MI model for the parameters $\mathbf{P}_0$ and the initial conditions $\mathbf{I}_0$.}

\Fig{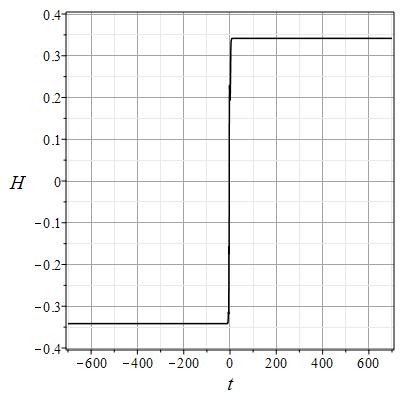}{8}{\label{H0} Evolution of the Hubble parameter $H(t)$ in the \MI model for the parameters $\mathbf{P}_0$ and the initial conditions $\mathbf{I}_0$.}

\Fig{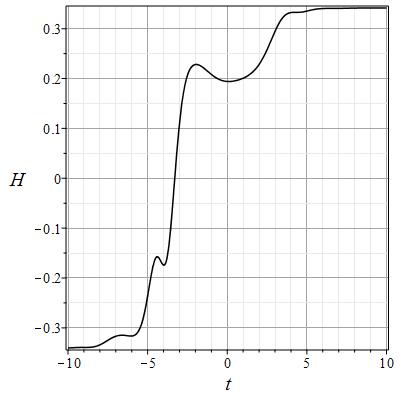}{8}{\label{H0_1} Evolution of the Hubble parameter $H(t)$ in the \MI model for the parameters $\mathbf{P}_0$ and the initial conditions $\mathbf{I}_0$ on a small scale.}

\Fig{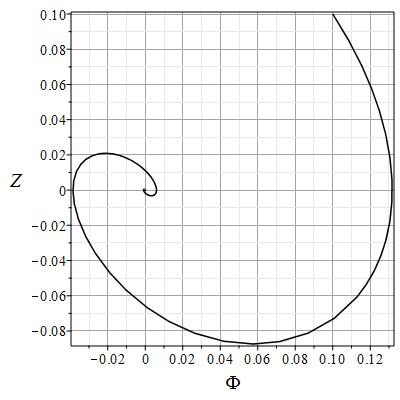}{8}{\label{FZ0} The phase trajectory of the model \MI in the canonical plane $\{\Phi,Z\}$ for the parameters $\mathbf{P}_0$ and the initial conditions $\mathbf{I}_0$.}

\Fig{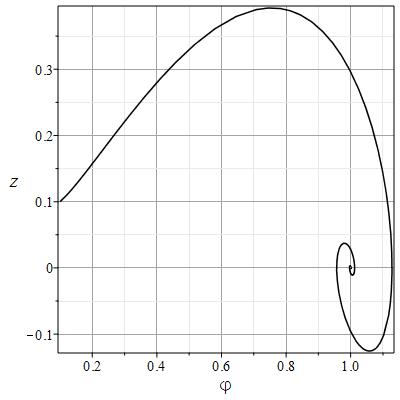}{8}{\label{f_z0}The phase trajectory of the model \MI in the phantom plane $\{\varphi, z\}$ for the parameters $\mathbf{P}_0$ and the initial conditions $\mathbf{I}_0$.}

\section{Example of oscillatory mode $\Lambda<0$}
In Fig. \ref{Ht_178_01-02} is shown a typical example of the oscillatory mode of the \MI model for the model parameters $\mathbf{P}_{11}=[[1,1,1,0,0.1],[1,0.5,1,0,0.1],-0.02]$ and $\mathbf{P}_{12}=[[1,1,1,0,0.2],[1,0.5,1,0,0.2],-0.02]$ and for the initial conditions $\mathbf{I}_1=[0.2,0.1,0.1,0.1,1]$.

\Fig{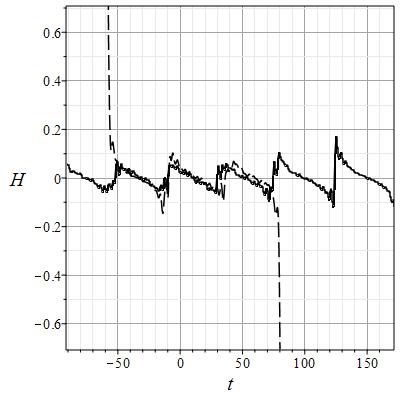}{8}{\label{Ht_178_01-02} Evolution of the Hubble parameter $H(t)$ in the \MI model: solid line $\mathbf{P}=\mathbf{P}_{11}$, dashed line -- $\mathbf{P}=\mathbf{P}_{12}$; initial conditions $I_1$.}

In Fig. \ref{Ht_178_01-02} one can see how an increase in the initial Fermi momentum by a factor of 2 for canonical and phantom $\zeta$ fermions led to a violation of the oscillatory regime of the cosmological model and to the finiteness of its lifetime $\Delta t=t_2-t_1\approx 80-(-50)=130$.

In Fig. \ref{Tr_178_2} and \ref{Tr_178_2f} are shown the phase trajectories of a dynamical system with the parameters $\mathbf{P}_{11}$.
\Fig{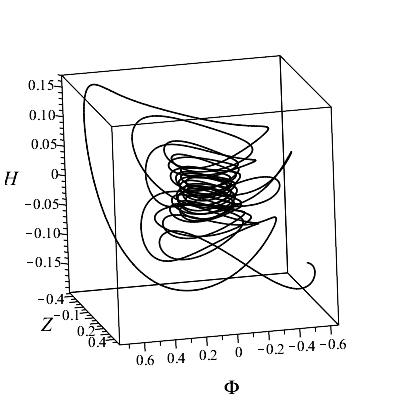}{7}{\label{Tr_178_2} The phase trajectory of the dynamic system of the \MI model in the canonical subspace $\mathbb{R}_3=\{\Phi,Z, H\}$ for the parameters $\mathbf{P}=\mathbf{P}_{11}$.}
\Fig{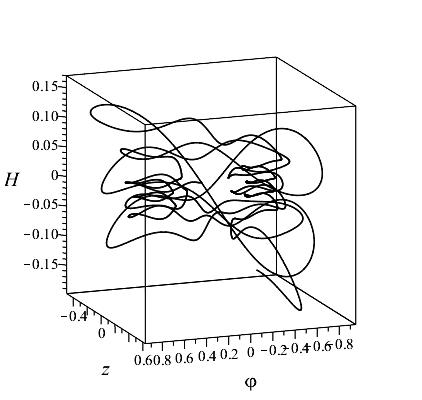}{7}{\label{Tr_178_2f} The phase trajectory of the dynamic system of the \MI model in the phantom subspace $\mathbb{R}_3=\{\varphi,z, H\}$ for the parameters $\mathbf{P}=\mathbf{P}_{11}$.}

It should be noted that these models are very close to the models with the vacuum asymmetric scalar doublet \cite{YuKokh_TMF}.

\section{Examples of cosmological models with a finite history}
As follows from the results of the section \ref{ssM1} in systems of scalar charged fermions, it is possible to achieve the cosmological singularity $a(t_1)\to0$ in the expansion phase of $H(t_1)\to+\infty$ and in the compression phase $a (t_2)\to0$, $H(t_2)\to-\infty$. Such a universe seems to exist for a limited time $t_2-t_1$. Note that in the incomplete \MOO model, due to the non-negativity of the Hubble parameter, the universe exists for infinite time.

We study in more detail the numerical models of such a process. Let us first find out how the values of the Fermi momentum affect the duration of the history of the universe $\Delta t$. In the set of parameters of the model $\mathbf{P}_{11}$, we fix all the parameters except for the Fermi pulses
\[\mathbf{P}_1(\pi_c,\pi_f)\equiv[[1,1,1,0,\pi_c],[1,0.5,1,0,\pi_f],-0.02],\]
while maintaining the initial conditions
\[\mathbf{I}_1=[0.2,0.1,0.1,0.1,1].\]
The Fig. \ref{xit_178_01} -- \ref{xit_178_019_c0} show graphs of the evolution of the scale function $ \ xi (t)$ depending on the value of the Fermi pulses in the range $0.1\div0.2$.

\Fig{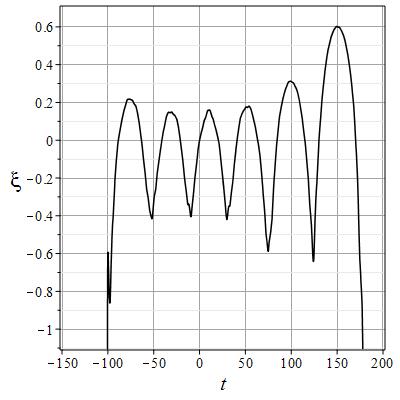}{8}{\label{xit_178_01} Evolution of the function $\xi(t)=\ln(a (t))$ for the model \MI for the parameters $\mathbf{P}_1(0.1,0.1)$.}

\Fig{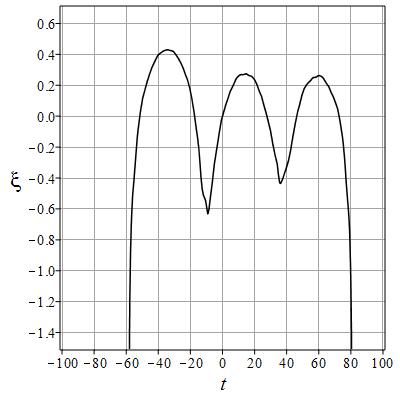}{8}{\label{xit_178_02} Evolution of the function $\xi(t)=\ln(a (t))$ for the model \MI for the parameters $\mathbf{P}_1(0.2,0.2)$.}

\Fig{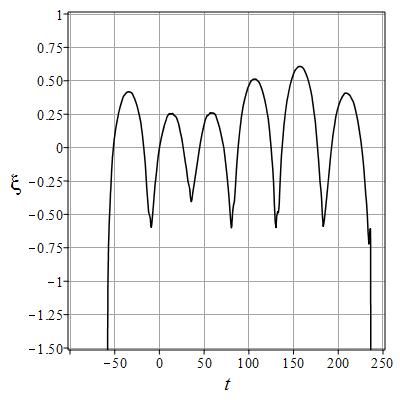}{8}{\label{xit_178_019} Evolution of the function $\xi(t)=\ln(a (t))$ for the model \MI for the parameters $\mathbf{P}_1 (0.19,0.19)$.}

At the same time, it can be noted that the longest lifetimes of the cosmological model correspond, first, to the values of the momentum $\pi_a=0.19$, and, secondly, in the presence of only one of the components of the Fermi system. Third, we can find that the lifetime of the cosmological model is approximately proportional to the number of$ n $ oscillations of the scale function $\xi(t)$, and the duration of each oscillation $\tau$ is approximately the same for all the presented cases $\tau\approx45$, except for one case (Fig. \ref{xit_178_019_f0}), in which it is equal to $\tau\approx35$.

\Fig{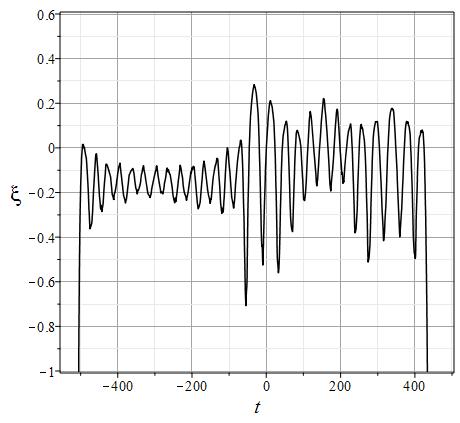}{8}{\label{xit_178_019_f0} Evolution of the function $\xi(t)=\ln(a (t))$ for the model \MI for the parameters $\mathbf{P}_1 (0.19,0)$.}

\Fig{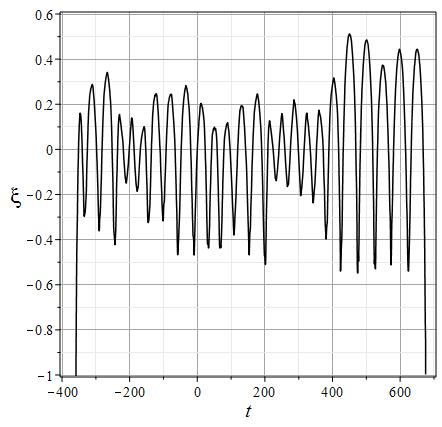}{8}{\label{xit_178_019_c0} Evolution of the function $\xi(t)=\ln(a (t))$ for the model \MI for the parameters $\mathbf{P}_1(0,0. 19)$.}

The lifetime of the model in the above examples varies within $\Delta t\approx 140\div 1020$. At the same time, it should be understood that in the absence of one of the charge carriers, the corresponding scalar field evolves as a minimally bound vacuum field in accordance with the corresponding initial conditions.

\section*{Conclusion}
Thus, firstly, we investigated the asymptotic and limiting properties of a cosmological model based on a two-component statistical system of degenerate scalar charged fermions interacting with an asymmetric scalar doublet. Secondly, we constructed a corresponding numerical model with the help of which we conducted a comparative analysis of this model with the previously studied incomplete models. Third, we have identified the possibility of oscillatory behavior of the model, as well as the possibility of the existence of models with finite lifetimes.

Summing up the results of this study, we will list its most important results.\\
\noindent$\bullet$ A closed mathematical model of a cosmological system consisting of 2 varieties of degenerate scalar charged fermions and an asymmetric pair of scalar fields, canonical $\Phi$ and phantom $\varphi$, with Higgs potential energy, describing a dynamical system in a 6-dimensional phase space $\mathbb{R}_6=\{\xi,H$, $\Phi,Z=\dot{\Phi}$, $\varphi,z=\dot{\varphi}\}$. The dynamical system is completely described by a normal autonomous system of ordinary differential equations with respect to the cosmological time $t$. Its behavior in a model with a cosmological term is determined by 11 parameters.\\
\noindent$\bullet$ The asymptotic and limiting properties of the models are investigated. It is shown that in cases that admit infinite expansion of the universe, its asymptotic properties at infinity are completely determined by vacuum scalar fields, i.e., mainly by late inflation. At the same time, the model also allows for finite histories of the universe, in these cases, the exit from the singular state and the entry into it occurs according to the asymptotics of the ultra-relativistic model. The ultimate histories of the universe are made possible precisely by the factor of the scalar charge of fermions.\\
\noindent$\bullet$ By numerical integration methods, the behavior of the studied cosmological model is compared both with the previously studied incomplete models of the cosmological evolution of charged degenerate fermions, and with the complete model based on the vacuum asymmetric scalar doublet. It is shown that in the first case, the behavior of the models coincides quite closely, except for the appearance of fluctuations in the Hubble parameter in the full model. In the second case, examples of the almost vacuum behavior of the model under study are shown. Examples of the oscillatory expansion mode characteristic of a model with vacuum fields and a negative cosmological constant are found. Thus, it is shown that the properties of the model coincide with the previously studied models in the limiting cases of the parameters.\\
\noindent$\bullet$ The above-mentioned cosmological models with a finite history between the ultra-relativistic exit from the singularity and the ultra-relativistic entry into the new singularity are studied by numerical simulation methods (Fig. \ref{xit_178_01} -- \ref{xit_178_019_c0}). In this case, the cosmological model in the interval between its beginning and end makes finite oscillations, the number of which is determined by the parameters of the model. The lifetime of the model is approximately proportional to the number of oscillations.

The preliminary analysis shows the need for a more detailed study of the formulated model and to clarify its dependence on the fundamental parameters of the system. The results of this study will be presented shortly.

\subsection*{Funding}

This paper has been supported by the Kazan Federal University Strategic Academic Leadership Program.

\end{document}